\documentclass{aa}  
\usepackage{graphicx}
\usepackage{natbib}
\usepackage{txfonts}
\bibpunct{(}{)}{;}{a}{}{,}
\begin{document}
   \title{The fraction of DA white dwarfs with
kilo-Gauss magnetic fields
\thanks{Based on observations made with ESO Telescopes at the La Silla or Paranal Observatories under programme ID 073.D-0356}
}

   \author{S. Jordan
          \inst{1}
          \and
          R. Aznar Cuadrado
          \inst{2}
        \and
        R. Napiwotzki
          \inst{3}
         \and
         H. M. Schmid          
          \inst{4}
        \and
         S. K. Solanki
          \inst{2}
          }

   \offprints{S. Jordan}

   \institute{Astronomisches Rechen-Institut, Zentrum f\"ur Astronomie der Universit\"at Heidelberg, M\"onchhofstr. 12-14,
       D-69120 Heidelberg, Germany\\
         \email{jordan@ari.uni-heidelberg.de}
         \and
         Max-Planck-Institut f\"{u}r Sonnensystemforschung,
       Max-Planck-Str. 2, D-37191 Katlenburg-Lindau, Germany\\
         \email{aznar@linmpi.mpg.de, solanki@linmpi.mpg.de}
      \and
        Centre for Astrophysics Research, University of Hertfordshire, Hatfield AL10 9AB, UK\\ 
         \email{rn@star.herts.ac.uk}
                \and
       Institut f\"ur Astronomie, ETH Z\"urich,
       CH-8092 Z\"{u}rich, Switzerland\\
         \email{schmid@astro.phys.ethz.ch}
   }

   \date{Received 2006; accepted}

 
  \abstract
   {
   Weak magnetic fields have been searched for on only a small number of white
dwarfs.
 Current estimates find that about 10\%\ of all white dwarfs have  fields in excess of 1\,MG;
   according to previous studies this number increases up to about 25\%\ in the kG regime.
}
   {Our aim is to improve on these statistics by a new 
   sample of ten white dwarfs in order
   to determine the ratio of magnetic to field-free white dwarfs. 
   }
   {Mean longitudinal magnetic fields strengths were determined by means of high-precision circular polarimetry of H$\beta$ and H$\gamma$ with the FORS1 spectrograph
    of the VLT ``Kueyen'' 8 m telescope. 
   }
   {In one of our objects (LTT\,7987), we detected a statistically
    significant (97\%\ confidence level) longitudinal magnetic
    field varying between ($-1\pm 0.5$)\,kG and ($+1\pm 0.5$)\,kG.
    This would be the weakest magnetic field ever found in a white dwarf,
    but systematic errors cannot completely
    be ruled out  at this level of accuracy.  We also observed the sdO star EC\,11481-2303 but could
    not detect a magnetic field.
   }
   {
VLT observations with uncertainties typically of 1000 G or less suggest that
15$-$20\% of WDs have kG fields.
 Together with previous investigations, the fraction of kG magnetic
    fields in white dwarfs amounts to about  11$-$15\%, which is 
    close to the current estimations for highly magnetic white dwarfs
    ($>$1\,MG). 
    }


   \maketitle
%

\section{Introduction}
The question how many white dwarfs are magnetic has been actively
debated in the recent years. In about 170 of the 5500 white dwarfs
listed in the online version of the Villanova White Dwarf Catalog 
(http://www.astronomy.villanova.edu/WDCatalog/index.html)
magnetic fields between 2\,kG and 1\,GG have been measured,
 corresponding to a fraction of
about 3\%\ \citep{McCook-Sion:99,wickramasinghe+ferrario00-1,vanlandinghametal05-1}. However, the spectra of only a few of the known white dwarfs
have been examined for the presence of a magnetic field in enough detail.
\citet{liebertetal03-1} and \citet{schmidtetal03-1} estimate that the 
true fraction of white dwarfs with magnetic fields in excess
of 2\,MG is expected to be at least $\approx 10\%$ and may be as high as 20\%.

Until recently, magnetic fields below  30 kG could not be detected, with
the exception of the very bright white dwarf 40 Eri B ($V=8.5$), in which
\citet{Fabrika-etal:03} found a magnetic field of 4\,kG.
However, by using the ESO VLT, we could push the detection limit
down to about 1\,kG in our first investigation of  12  DA white dwarfs
with $11<V<14$ \citep{Aznar-etal:04}. In 3 objects of this sample,
we detected magnetic fields between 2\,kG and 7\,kG on a $5\sigma$ 
confidence level. Therefore, we concluded that the fraction of white dwarfts
with kG magnetic fields is
about 25\%. 

For one of our cases, LP\,672$-$001 (\object{WD\,1105$-$048}),  \citet{Valyavin-etal:06}
confirm the presence of a kG magnetic field by measuring circular polarisation at H$\alpha$ using the 
6m of the Special Astrophysical Observatory.
On the other hand, none of the other 5 bright white dwarfs of their sample showed any significant signature of a magnetic field. 
They detected a magnetic field of up to 10\,kG 
in the hot subdwarf (spectral type sdO) \object{Feige\,34}, confirming the
detection of magnetic fields in both types of white dwarf
progenitors, hot subdwarfs \citep{otooleetal05-2} and
central stars of planetary nebulae \citep{jordanetal05-1}.

With this new investigation, we increase the sample
of white dwarfs that is checked for kG magnetic fields by means
of circular polarisation by eleven objects (one turned out to be a high-metallicity
sdO, \citet{Stys-etal}). This should allow us to
provide a much better estimate of the incidence of low magnetic fields 
in white dwarfs.

All tables have been organised in exactly the same way as 
in  \citeauthor{Aznar-etal:04} (\citeyear{Aznar-etal:04}, Paper\,I); the figures show the same
spectral regions as in Paper\,I but have been plotted in a more compact way.

\section{Observations and data reduction}
The spectropolarimetric data of our new sample of  ten bright normal
DA white dwarfs plus one high-metallicity sdO star were obtained in service mode between May 5 and August
4, 2004, with the FORS1 spectrograph at the 8\,m UT2 (``Kueyen'') of the
VLT. The setup was exactly the same as described in Paper\,I.
The spectra and circular polarimetric data covered the wavelength region
between 3600\,\AA\ and 6000\,\AA\ with a spectral resolution of
4.5\,\AA. A higher spectral resolution would not provide a higher
sensitivity since the accuracy is basically limited by the signal-to-noise
ratio allone. 
The exposures were split into
a sequence of exposures to avoid saturation; after every second observation the retarder 
plate was rotated from $\alpha=-45^\circ$ to $\alpha=+45^\circ$ and back
in order to suppress spurious signals in the degree of circular polarisation
(calculated from the ratio of the Stokes parameters $V$ and $I$). 
All stars were observed in two or three different nights in order to
detect the presence of possible variations in $V/I$ due to the rotation of the stars.

As in the case of the sample of Paper\,I, all objects 
in our new sample of DA white dwarfs have
been previously observed in the course of the  SPY survey \citep{Napi-etal:03},
a radial velocity search for close binary systems  composed of two white
dwarfs.
We checked all  candidates for spectral
peculiarities and magnetic fields strong enough to be detected in intensity
spectra taken with the  high-resolution Echelle spectrograph UVES at the
Kueyen (UT2) of the VLT.
None of our programme stars showed any sign of Zeeman splitting in the SPY
SURVEY, i.e., indicating that any possible magnetic field must be below a level of about 20\,kG.

The data reduction and calculation of the observed circular polarisation
is described in detail in Paper\,I. Special care was taken
to avoid errors from  changes in the sky transparency,
atmospheric scintillation, and various instrumental effects. The 
wavelength calibration was
made for each observing date separately, and no spurious signals were detected during the calibration process.

Details of our eleven sample stars and of our observations are listed in 
Table\,\ref{t:obs}. 
The provided $\alpha$ and $\delta$
coordinates refer to epoch 2000 as measured in the course of the SPY project
\citep[see][]{Koester-etal:01}.
Spectral types, $T_{\rm sp}$, and measured V magnitudes were taken from the
catalogue of \citet{McCook-Sion:99}.
EC\,11481-2303 was classified as an DAO white dwarf by \citet{Kilkenny-etal:97}.

\section{Determination of magnetic fields}
As  discussed in Paper\,I, the theoretical $V/I$ profile
for a given mean longitudinal magnetic field $\langle B_z\rangle$ 
(expressed in Gauss) below
about 10\,kG is given by  the weak-field approximation
\citep[e.g.\,][]{Angel-Landstreet:70,Landi-Landi:73} without any loss in
accuracy:
\begin{equation}
\frac{V}{I} = -g_{\rm eff} \ensuremath{C_z}\lambda^{2}\frac{1}{I}
\frac{\partial I}{\partial \lambda}
\ensuremath{\langle\large B_z\large\rangle}\;,
\label{e:lf}
\end{equation}
where $g_{\rm eff}$ is the effective Land\'{e} factor
\citep[= 1 for all hydrogen lines of any series,][]{Casini-Landi:94},
$\lambda$ is the wavelength expressed in \AA, 
and the constant
$C_z=e/(4\pi m_ec^2)$ $(\simeq 4.67 \times10^{-13}\,{\rm G}^{-1} \mbox{\AA}^{-1})$.

We again performed a $\chi^2$-minimisation procedure  to find out
which mean longitudinal magnetic field strength fits the 
observed data best in wavelength  intervals  of $\pm$20\,\AA\ around H$\beta$ 
and H$\gamma$.
Since the  error
of the magnetic field determination increases for the higher series number,
we based our investigation only on these two Balmer  lines.

The resulting best-fit values for the magnetic field strengths from 
the individual lines and their statistical $1\sigma$ errors are listed in
Table\,\ref{t:mf} for each observation. 
$B({\sigma})$ provides the magnetic field in units of
the $\sigma$ level. Detections exceeding the $2\sigma$ levels are given in
bold. $\sigma(V)$ is the standard deviation of the observed ($V/I$)-spectrum
obtained in the region 4500--4700\,\AA. Lower limits on the detectability of
the magnetic field from the line polarisation peaks, calculated at the
$1\sigma$ level of the noise, are given in the last two columns. Multiple
observations that were averaged prior to analysis are labeled {\sc average}.
We also provid the weighted means
$B_z=(B_{z,\gamma} w_\gamma+B_{z,\beta} w_\beta)/(w_\gamma+w_\beta)$ where
$w_i=1/\sigma_i^2$ ($i=\gamma,\beta$). The probable error is given by
$\sigma=(\sum w_i)^{-1/2}$.

Moreover, we expressed the resulting mean magnetic field in multiples
of the $1\sigma$ error in order to judge the significance of the 
magnetic field determination. For a comparison we note that
for each of the three stars
 with significant
magnetic fields from the \citet{Aznar-etal:04} sample, at least one observation
exceeded the $5\sigma$ level.

\begin{figure*}
   \centering
  \includegraphics[width=0.8\textwidth]{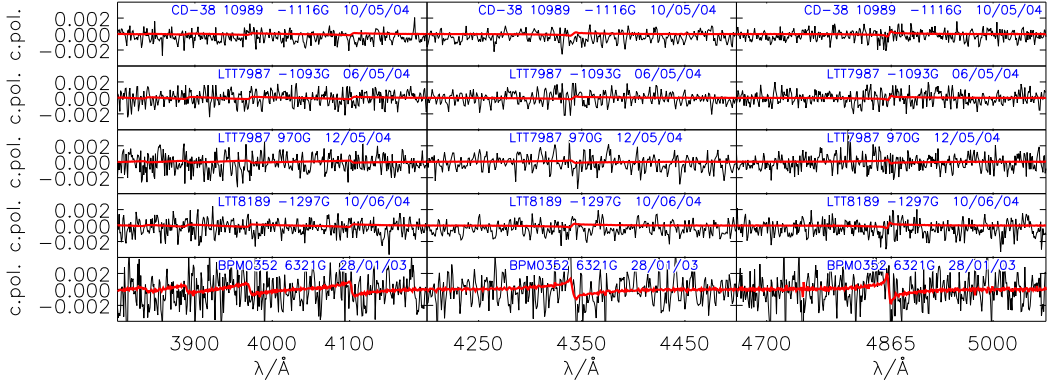}
   \caption{Circular polarisation spectra $V/I$ (thin black lines)
around H$\beta$ (4861\AA, right), H$\gamma$ (4340\,\AA, middle), and close
to the Balmer threshold (left)
for those observations, where the $2\sigma$ level is exceeded.
The thick lines (red in the online version)
 represent the circular 
polarisation predicted by the low-field approximation (Eq.\,\ref{e:lf}), using
the values for the longitudinal magnetic field from the best fit to the
H$\beta$ and  H$\gamma$ regions. For comparison we also plotted the
result for \object{BPM\,0352} with $B=6321$\,G (lowest panel),
 which was analysed by \citet{Aznar-etal:04}. The plots of all other observations are
shown in Fig.\,\ref{wd73a2} of the online version.}
              \label{wd73a}%
    \end{figure*}

In Table\,\ref{t:mf} we also calculated lower limits on the detectability
of the magnetic fields from the line polarisation peaks, which guide the
eye when judging the confidence of the fits from plots of the 
circular polarisation (Fig.\ref{wd73a} and online Fig.\ref{wd73a2}). Note, however,
that a significant contribution to the fit originates not only from the
narrow peaks but also from the full 40\,\AA\ interval around the Balmer lines.
Note, however, that in the three magnetic cases of our first
study, the S-wave circular polarisation signature reaches the $2\sigma$
level of the noise. 
For each object we also added up all measurements according to 
formula 3 in Paper\,I. The results for these added observations
were labeled as {\sc average}. 

Our fitting procedure was validated with extensive numerical simulations
using a large sample of 1000 artificial noisy polarisation spectra
\citep{Aznar-etal:04}. It was found that at our noise level 
$\sigma(V)$ (also listed in Table\,\ref{t:mf}) 
of some $10^{-3}I_{\rm c}$ ($I_{\rm c}$ being the continuum
intensity), kG fields can reliably be detected.

   \begin{figure}
   \centering
   \includegraphics[width=6.8cm,angle=270]{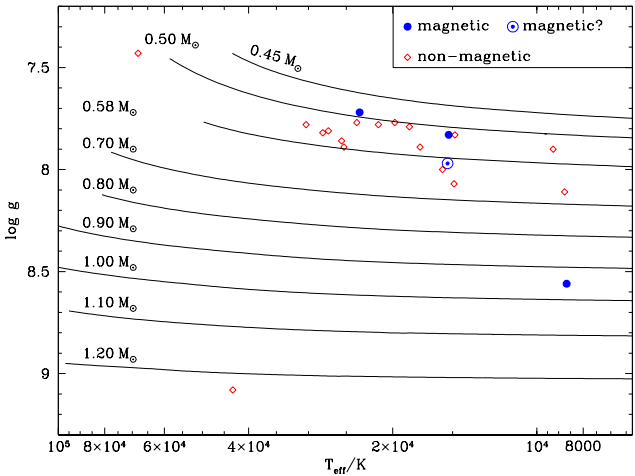}
   \caption{
Temperature-gravity diagram with the positions of white
dwarfs analysed in Paper~I and in the current paper. Theoretical
cooling sequences from \citet{Benvenuto+Althaus:99} are plotted for
various white dwarf masses. White dwarfs with and without detected
magnetic fields are indicated by different symbols. The double-lined
binary \object{WD\,0135$-$052} from Paper~I is not shown in this plot.
}
              \label{f:hr_mag}%
    \end{figure}

\begin{table*}
\begin{center}
\caption{Details of the VLT observations.}
\label{t:obs}
\begin{tabular}[c]{llrrrcrrl}
\hline
\hline
\multicolumn{1}{c}{Target} & \multicolumn{1}{c}{Alias} &
\multicolumn{1}{c}{$\alpha$} & \multicolumn{1}{c}{$\delta$} &
\multicolumn{1}{c}{V} & \multicolumn{1}{c}{HJD} &
\multicolumn{1}{c}{$t_{\rm exp}$} & \multicolumn{1}{c}{$n$} &
\multicolumn{1}{c}{$T_{sp}$} \\
\multicolumn{1}{c}{} & \multicolumn{1}{c}{} & \multicolumn{1}{c}{} &
\multicolumn{1}{c}{} & \multicolumn{1}{c}{(mag)} &
\multicolumn{1}{c}{(+2453000)} &
\multicolumn{1}{c}{(s)} & \multicolumn{1}{c}{} & \multicolumn{1}{c}{} \\
\hline
\object{WD\,1148$-$230} &
\object{EC\,11481$-$2303}& 11 50 38.85 & $-$23 20 34.8 & 11.76 & 134.096
& 150 & 10 & sdO  \\
               &              &             &               &       & 144.097
& 150 & ~8 &    \\
\object{WD\,1202$-$232} & \object{EC\,12028$-$2316} & 12 05 26.80  & $-$23 33 14.0   & 12.79 & 144.132
& 500 & ~4 & DA \\
               &              &             &               &       & 151.983
& 500 & ~4 &    \\
\object{WD\,1327$-$083} & \object{G\,14$-$058}     & 13 30 13.58 & $-$08 34 30.2 & 12.31 & 151.519 & 290 & ~6 & DA3.5 \\
              &               &            &             &       & 153.554 & 285 & ~6 &     \\
\object{WD\,1620$-$391}   & \object{CD$-$38\,10980} & 16 23 33.84 & $-$39 13 46.2 & 11.00  & 136.285
& ~73 & 14 & DA \\
               &              &             &               &       & 143.323
& ~73 & 14 &    \\
               &              &             &               &       & 151.054
& ~73 & 14 &    \\
\object{WD\,1845$+$019} & \object{Lan\,18}     & 18 47 39.09 & $+$01 57 33.8 & 12.95 & 131.878 & 350 & ~6 & DA1.5 \\
              &               &            &             &       & 136.873 & 285 & ~6 &     \\
\object{WD\,1919$+$145} & \object{GD\,219}     & 19 21 40.51  & $+$14 40 40.5& 12.94 & 132.808 & 350 & ~6 & DA5 \\
              &               &            &             &       & 136.834 & 350 & ~6 &     \\
\object{WD\,2007$-$303} & \object{LTT\,7987}    & 20 10 56.82  & $-$30 13 06.7& 12.18 & 132.848 & 300 & 12 & DA4 \\
              &               &            &             &       & 138.858 & 300 & ~6 &     \\
\object{WD\,2014$-$575} & \object{RE\,2018$-$572} & 20 18 54.88 & $-$57 21 33.8& 13.00 & 140.842 & 350 & 6 & DA2 \\
              &               &            &             &       & 184.757 & 350 & ~2 &     \\
              &               &            &             &       & 185.591 & 350 & ~6 &     \\
\object{WD\,2039$-$202} & \object{LTT\,8189}    & 20 42 34.64  & $-$20 04 35.6& 12.33 & 143.847 & 300 & ~6 & DA2.5 \\
              &               &            &             &       & 167.879 & 300 & ~6 &     \\
\object{WD\,2149$+$021} & \object{G\,93$-$048}  & 21 52 25.43   & $+$02 23 17.8 & 12.72 &183.762 & 348 & ~6 & DA3 \\
              &               &            &             &       & 196.829 & 348 & ~6 &     \\
              &               &            &             &       & 222.684 & 348 & ~6 &     \\
\object{WD\,2211$-$495} & \object{RE\,2214$-$491} & 22 14 11.93 & $-$49 19 27.1& 11.70 & 140.885 & 161 & 10 & DA1 \\
              &               &            &             &       & 185.730 & 161 & 10 &     \\
\hline
\end{tabular}
\end{center}
\end{table*}

\begin{table*}
\begin{center}
\caption{Magnetic fields derived from the H$\gamma$ and H$\beta$ lines for our
sample of white dwarfs.
}
\label{t:mf}
\begin{tabular}[c]{lccrrrrrr}
\hline
\hline
\multicolumn{1}{c}{Target} & \multicolumn{1}{c}{Date} &
\multicolumn{1}{c}{$\sigma(V)$} & \multicolumn{2}{c}{$B$(G)} &
\multicolumn{1}{c}{$B({\rm G})$} & \multicolumn{1}{c}{$B({\sigma})$} &
\multicolumn{2}{c}{B(G) at $1\sigma$}\\
\multicolumn{1}{c}{} & \multicolumn{1}{c}{} &
\multicolumn{1}{c}{(10$^{-3}I_{\rm c}$)} &
\multicolumn{1}{c}{H$\gamma$} & \multicolumn{1}{c}{H$\beta$} &
\multicolumn{1}{c}{H$\gamma, \beta$} & \multicolumn{1}{c}{H$\gamma, \beta$} &
\multicolumn{1}{c}{~~~H$\gamma$} & \multicolumn{1}{c}{~~~H$\beta$} \\
\hline
\object{WD\,11481$-$2303}&08/05/04     & 0.7 &  $-520\pm 655$&    $-980\pm 590$&    $-774\pm438  $ & 1.76 & $-$1970 &  $-$1480 \\
            &18/05/04     & 1.0 &  $-30\pm1325$&    $20\pm1095$&    $0\pm844  $ & 0.00 & 3060 &  2310 \\
            &{\sc average} & 0.6 &  $-490\pm625$&     $-860\pm500$&    $-716\pm390  $ & 1.83 & $-$1650 &  $-$1260 \\[0.08cm]
\object{WD\,1202-232}&18/05/04     & 1.3 &  $1280\pm 865$&    $260\pm 940$&    $812\pm636  $ & 1.28 & 1560 &  1160 \\
            &25/05/04     & 1.2 &  $660\pm550$&    $-370\pm325$&   $-103\pm280  $ & 0.37 & 1260 & 960 \\
            &{\sc average} & 0.9 &  $-200\pm260$&     $850\pm425$&    $85\pm221  $ & 0.39 & 1040 &  770 \\[0.08cm]
\object{WD\,1327-083}&25/05/04     & 0.8 &  $300\pm1010$&    $-320\pm1080$&    $10\pm737  $ & 0.01 & 2110 &  $-$1690 \\
            &27/05/04     & 1.1 &  $-2800\pm740$&     $2790\pm835$&   $-340\pm553$ & 0.61 & $-$2700 &  2220 \\
            &{\sc average} & 0.7 &  $-1230\pm555$&     $1410\pm520$&    $175\pm379$ & 0.46 & $-$1630 &  1330 \\[0.08cm]
\object{WD\,1620-391}&10/05/04     & 0.6 &  $150\pm 785$&    $-1580\pm 475$&  {\bf $-$1116$\pm$ 406}  & 2.75 &-1900 & $-$1330 \\
            17/05/04     & 0.7 &  $-2390\pm1220$&    $-20\pm735$&   $-651\pm629  $ & 1.03 &-2150 & $-$1490 \\
            &25/05/04     & 0.6 &  $120\pm640$&    $-770\pm595$&   $-357\pm435  $ & 0.82 & $-$1990 & $-$1440 \\
            &{\sc average} & 0.4 &  $-500\pm480$&     $-920\pm365$&  {\bf  $-$766$\pm$290} & 2.63 & 1310 & 910 \\[0.08cm]
\object{WD\,1845+019}&05/05/04     & 1.1 &  $-340\pm1410$&    $-150\pm1175$&  $-227\pm902  $ & 0.25 & $-$4660 &  $-$3710 \\
            &10/05/04     & 0.9 &  $-30\pm1145$&    $6000\pm2390$&  $1095\pm1032 $ & 1.06 & $-$3930 &  2930 \\
            &{\sc average} & 0.8 &  $-130\pm815$&      $500\pm670$&    $245\pm517  $ & 0.47 & $-$3310 &  2510 \\[0.08cm]
\object{WD\,1919+145}&06/05/04     & 1.6 &  $1240\pm1080$&   $-1440\pm1220$&    $62\pm808  $ & 0.08 & 4590 &  $-$3610 \\
            &10/05/04     & 1.2 &  $ 930\pm1290$&     $110\pm990$&    $413\pm785  $ & 0.53 & 3340 &  2650 \\
            &{\sc average} & 1.0 &  $1180\pm870$&      $-500\pm815$&    $285\pm594$ & 0.48  & 2760 &  $-$2150 \\[0.08cm]
\object{WD\,2007$-$303}&06/05/04     & 0.7 &  $-1460\pm1270$&   $-1040\pm485$&    {\bf $-$1093$\pm$ 453} & 2.41 & $-$1780 &  $-$1280 \\
            &12/05/04     & 0.8 &  $1540\pm780$&     $610\pm620$&    {\bf 970$\pm$485}  & 2.00 & 2270 &  1640 \\
            &{\sc average} & 0.5 &  $120\pm495$&      $-390\pm375$&    $-204\pm298$ & 0.67 & 1460 &  $-$1090 \\[0.08cm]
\object{WD\,2014-575}&14/05/04     & 1.6 &  $2410\pm1735$&    $1240\pm1260$&  $1643\pm1019$ & 1.61 & 5340 &  3950 \\
            &27/06/04     & 3.4 &  $310\pm4295$&    $1160\pm3340$&   $839\pm2636 $ & 0.32 & 11680  &  7720 \\
            &28/06/04     & 1.8 &  $2820\pm2060$&   $-4470\pm1455$& $-2043\pm1188 $ & 1.71 & 6330 &  $-$4130 \\
            &{\sc average} & 1.3 &  $2380\pm1205$&   $-1120\pm895$&    $124\pm718  $ & 0.17 & 3740 &  $-$2680 \\[0.08cm]
\object{WD\,2039-202}&17/05/04     & 1.4 &  $-2670\pm1595$&      $40\pm965$&   $-686\pm825  $ & 0.83 & $-$3650 &  2780 \\
            &10/06/04     & 0.8 &  $-780\pm730$&    $-1800\pm720$&  {\bf $-$1297$\pm$512} & 2.53 &$-$2370 &  $-$1670 \\
            &{\sc average} & 0.6 &  $-1240\pm655$&    $-1290\pm535$& {\bf $-$1269$\pm$414}   & 3.06 & $-$1810 &  $-$1300 \\[0.08cm]
\object{WD\,2149+021}&26/06/04     & 2.1 &  $340\pm1060$&     $730\pm930$&    $560\pm699  $ & 0.80 & 6570 &  4450 \\
            &09/07/04     & 1.1 &  $-530\pm945$&    $-1690\pm890$&  $-1144\pm647  $ & 1.78 & $-$3620 &  $-$2420 \\
            &04/08/04     & 0.8 &  $-1300\pm985$&      $350\pm705$&   $-208\pm573  $ & 0.36 & $-$2330 &  1520 \\
            &{\sc average} & 0.9 &  $-600\pm555$&     $-130\pm490$&   $-335\pm367  $ & 0.91 & $-$1730 &  $-$1150 \\[0.08cm]
\object{WD\,2211-495}&14/05/04      & 0.7 &  $110\pm1655$&   $-1940\pm1060$& $-1343\pm892  $ & 1.51 & 7570 &  $-$4150 \\
           &28/06/04      & 0.7 &  $-190\pm1795$&     $640\pm1190$&   $386\pm991  $ & 0.39 & 8570 &  4470 \\
           &{\sc average}  & 0.5 &  $-390\pm1155$&    $-900\pm795$&   $-736\pm654  $ & 1.12 & $-$5830 &  $-$3050 \\[0.08cm]
\hline
\end{tabular}
\end{center}
\end{table*}
\begin{table*}
\begin{center}
\caption{Fitted parameters of the white dwarfs and supplementary data from
literature.
}
\label{t:lit}
\begin{tabular}[c]{l|rlcrr|rl|cc}
\hline
\hline
   && &&&&\multicolumn{4}{c}{literature} \\
WD & T$_{\rm eff}$ & $\log g$ & $M$ & $d$(spec) &$t_{\mathrm{cool}}$
   & $T_{\rm eff}$
   &$\log g$ & $v\sin i$ & $d$(trig)\\
   &(kK) && (${\rm M}_\odot$) &(pc) &(Myr)
   &(kK) && (km s$^{-1}$) &(pc)\\
\hline
\object{WD\,1202$-$232} & 8.75&8.11&0.663  &9.9  &1052  &  8.62    & 8.00$^7$ &$<6$$^7$       \\
\object{WD\,1327$-$083} &14.83&7.83&0.518  &18.1 &157 &14.41 &7.85$^1$ & & 
$18.0\pm 1$$^{2}$ \\
\object{WD1620$-$391}   &25.29&7.89&0.576  &14.9 &18 & 24.25 & 8.05$^4$   & $<8$$^4$     &    $12.8\pm 0.4$$^2$       \\
\object{WD\,1845$+$019} &30.33&7.78&0.536  &47.4 &8 &30.35 &7.83$^1$ & & \\
\object{WD\,1919$+$145} &14.88&8.07&0.652  &21.0 &236 &14.60 
&8.09$^{3}$&$<6$$^{4}$ &$20\pm3$$^{5}$ \\
\object{WD\,2007$-$303} &15.36&7.97&0.595  &16.0 &178 &15.15 &7.86$^1$ 
&$<7$$^{4}$  & $15\pm1$$^{2}$\\
\object{WD\,2014$-$575} &27.99&7.82&0.548  &57.8 &11 &28.37 &7.87$^1$ 
&$<14$$^{4}$ & \\
\object{WD\,2039$-$202} &19.79&7.77&0.502  &24.2 &46 &20.41 &7.84$^1$ 
&$<15$$^{4}$  & $21\pm2$$^{2}$\\
\object{WD\,2149$+$021} &17.53&7.89&0.557  &24.2 &96 &17.65 &7.99$^1$ & 
&$25\pm3$$^{2}$ \\
\object{WD\,2211$-$495} &68.11&7.43&0.518  &67.3 &0.1 &66.50 &7.52$^6$ & &  \\
\hline
\end{tabular}
\\
\end{center}
\footnotesize{References:
$^1$\citet{Bragaglia-etal:95};
$^{2}$\citet{Perryman-etal:97};
$^{3}$\citet{Koester-etal:98};
$^{4}$\citet{Karl-etal:05};
$^{5}$\citet{vanAltena-etal:95};
$^{6}$\citet{Finley-etal:97};
$^{7}$\citet{Berger-etal:05}.}
\end{table*}

\section{Results}
\object{EC\,11481$-$2303} (\object{WD1148$-$230}) is a high-gravity pre-white dwarf
of spectral type sdO, which we  
disregard n our statistics on white dwarfs, but whose measurement is intersting on its own.
The nature of this star with $T_{\rm eff}=42\,000$\,K and $\log g=5.8$ has been revealed by
\citet{Stys-etal}.

As can be seen from Table\,\ref{t:mf}, Fig.\,\ref{wd73a}, and Fig.\,\ref{wd73a2} (online only),
 none of the measurements of the circular polarisations
reached the same level of confidence as the three magnetic objects
found in the first sample (Paper\,I). The highest level of
confidence was achieved by \object{LTT\,7987} (\object{WD\,2007$-$303}) where a
$2.4$ and $2.0\sigma$ level was reached for the two respective observations.
The corresponding mean longitudinal field strengths were
$-1093\pm 453$\,G and $970\pm 485$\,G. Single observations  of 
\object{CD$-$38 10980} (\object{WD162$-$391}) resulted in $-1116\pm 406$\,G and
\object{LTT\,8189} (\object{WD\,2039$-$202})  in $-1297\pm 512$\,G, which
corresponds to $2.8\sigma$ and $2.5\sigma$, respectively. 

In Paper\,I we  disregarded the case of LHS\,1270, 
for which a magnetic field of $654\pm320$\,G was found. Since their
sample consisted of 22 single observations, statistically 
one would expect this $2\sigma$ observation even if all
the stars have no magnetic field. 

Our new sample consists of 23 single observations  of white dwarfs and
two observations of a metal-rich sdO. With the same
argument we would statistically expect only one observation to exceed the
$2\sigma$ level. Therefore, one can assume
that between none and {three} of the observations
may actually be a real observation. 

With two
observations exceeding $2\sigma$, \object{LTT\,7987} (\object{WD\,2007$-$303})
would be the most convincing  for   a positive detection.
The probability that two independent
and uncorrelated observations 
of a single star have that level of confidence can be estimated
in the following way:
The likelihood that an observation exceeds 2$\sigma$ is 4.6\%; therefore,
the chance that at least one observation  of the white dwarfs exceeds 2$\sigma$ is
$(1-0.954^{23})=66.1$\%. Then the probability that the same star has a
second observation exceeding 2$\sigma$ is $0.661\cdot 0.046=3.0$\%.
Therefore, 
from a purely statistical point of view, we must regard this detection
as significant (with 97\%\ confidence).

However, we have to take into account that the measured magnetic field
strengths would be only about 1\,kG, which is 2 to 6 times smaller than the
positive detections from the sample of Paper\,I. 
At this level we cannot fully exclude  systematic errors 
from the  limitation of our low-field approximation (assuming a single
magnetic strength rather than a distribution) or from instrumental
polarisation.
However, \citet{Bagnulo-etal:06} have shown that none of their observed
non-magnetic A stars -- observed with the same instrument --  showed
circular polarization hinting at magnetic fields larger than 400\,G. 
This would mean that the systematic errors are well below 1\,kG.

On the other hand,  none of the polarisation peaks of LTT\,7987 in Fig.\,\ref{wd73a}
exceeds the noise level, differently from the three detections 
\object{BPM\,03523}, \object{LP\,672$-$001}, and \object{L\,362$-$434} in \citet{Aznar-etal:04}. Therefore,  a $\chi$-by-eye analysis would not confirm our
detection but, as we pointed out in Paper\,I, this would be 
very misleading.

Both measurements of the sdO star EC\,11481-2303 are below the $2\sigma$ level.
This is interesting in itself and confirms the finding by \citet{otooleetal05-2} that
there is no correlation between the metallicity and the presence of a magnetic
field with kG strength.
Therefore, we conclude that although the measured magnetic field in LTT\,7987
is statistically  
significant, further observations are needed to establish their reality.

In order to find out whether white dwarfs with and without kG magnetic
fields differ in mass or age, we  computed masses and cooling ages
from the fundamental atmospheric parameters  temperature and gravity.
The values of $T_{\rm eff}$ and $\log g$ were derived from  a model atmosphere analysis of the
high signal-to-noise spectra (see Fig.\,\ref{f:spectra} in the online material). The observed line profiles are fitted with theoretical spectra from a
large grid of NLTE spectra calculated with the NLTE code developed by
\citet{Werner:86}.  The
four coolest white dwarfs of our sample (\object{WD\,1327$-$083},
\object{WD\,1919$+$145}, and \object{WD\,2007$-$303})
were analysed with a grid of LTE model
spectra computed by D.~Koester for the analysis of DA white dwarfs
\citep[see e.g.\,][] {Finley-etal:97}, which is more reliable below 17000\,K,
where convection and collision-induced
absorption by hydrogen quasi-molecules play a role. 

Table\,\ref{t:lit} lists the results of the model atmosphere analyses.
From these we determined spectroscopic distances,  as well
as masses and cooling ages computed from a comparison of parameters
derived from the fit with the grid of white dwarf cooling sequences by
\citet{Benvenuto+Althaus:99}, for an envelope hydrogen mass of
$10^{-4}M_{\mathrm{WD}}$. A temperature-gravity diagram with the positions 
of white dwarfs analysed in Paper~I and in the current paper is shown in 
Fig.\,\ref{f:hr_mag}.
 In Table\,\ref{t:lit} we  also provid data collected from the literature (atmospheric parameters, rotational velocities, and trigonometic parallaxes)
when available.

\section{Conclusion}
While we detected magnetic fields in  3 out of the  12 programme stars 
in our first investigation,
we found at most (if at all) one object in our new sample
of 10 DA white dwarfs. 
Putting both samples together, we arrive at a fraction of 
14$-$18\%\  of kG magnetism in white dwarfs; the lower value is obtained
assuming that
LTT\,7987 is not magnetic. However, if confirmed, LTT\,7987 would 
have the lowest magnetic field (1\,kG) ever detected in a white dwarf.

Recently, \citet{Valyavin-etal:06} also performed a search for 
circular polarisation in white dwarfs. They confirmed our detection
\citep{Aznar-etal:04} 
of a varying longitudinal magnetic field in \object{LP\,672$-$001} (\object{WD\,1105$-$048}):
they measured field strengths between $-7.9\pm 2.6$\,kG to $0.1\pm2.7$\,kG,
compared to our values of $-4.0\pm 0.7$\,kG to $-2.1\pm0.4$\,kG. 
However, they did not discover any significant magnetic field in their 
five other programme stars. If we combine their results with our's,
the fraction of kG magnetic fields in DA white dwarfs amounts to
15\%\ (4/(12+10+5)) or 11\%\ (3/(12+10+5)), if we disregard the detection in 
LTT\,7987. However, it is  problematic to merge both samples, because
the signal-to-noise ratio of our VLT measurements is  much higher than
the observations with the 6\,m telescope of the Special Astrophysical Observatory. Since our 
uncertainties are on the average 2$-$3 times smaller (partly also due to the 
fact that \citet{Valyavin-etal:06} only used H$\alpha$),
we must put a higher statistical weight on our sample
with a fraction of 11\%\ to 15\%\ of magnetic to field-free 
(i.e. below detection limit) white dwarfs.

While one of the white dwarfs with a magnetic field (L\,362$-$81) 
had an exceptionally high mass (0.95\,$M_\odot$), all objects with
kG magnetic fields have usual white dwarf masses in the range
0.5 to 0.6\,$M_\odot$. Therefore, a trend towards higher masses
in magnetic white dwarfs compared to non-magnetic ones \citep{Liebert:88}
does not seem to exist for magnetic white dwarfs with kG fields. 
This would be consistent with the idea that the high-magnetic-field  white 
dwarfs have a higher-than-average mass because they have more massive
progenitors. Therefore, it is probable that the white dwarfs with 
kG magnetic fields stem from a low-mass progeny on the main sequence 
(e.g.\, F stars) as  speculated by \citet{wickramasinghe+ferrario05-1}.

\begin{acknowledgements}
We acknowledge the use of LTE model spectra computed
by D. Koester, Kiel. We thank the staff of the ESO VLT for carrying out the service
observations. We thank U. Bastian, Heidelberg, for suggestions concerning the
correct application of statistics. 
G. Mathys and S. Bagnulo, both at ESO, 
have contributed to our project with valuable discussions.
\end{acknowledgements}

\Online
\begin{appendix}
\section{Figures in the electronic version}
Two figures are only available in the electronic online version of the paper.
   \begin{figure*}
   \centering
   \includegraphics[width=0.8\textwidth]{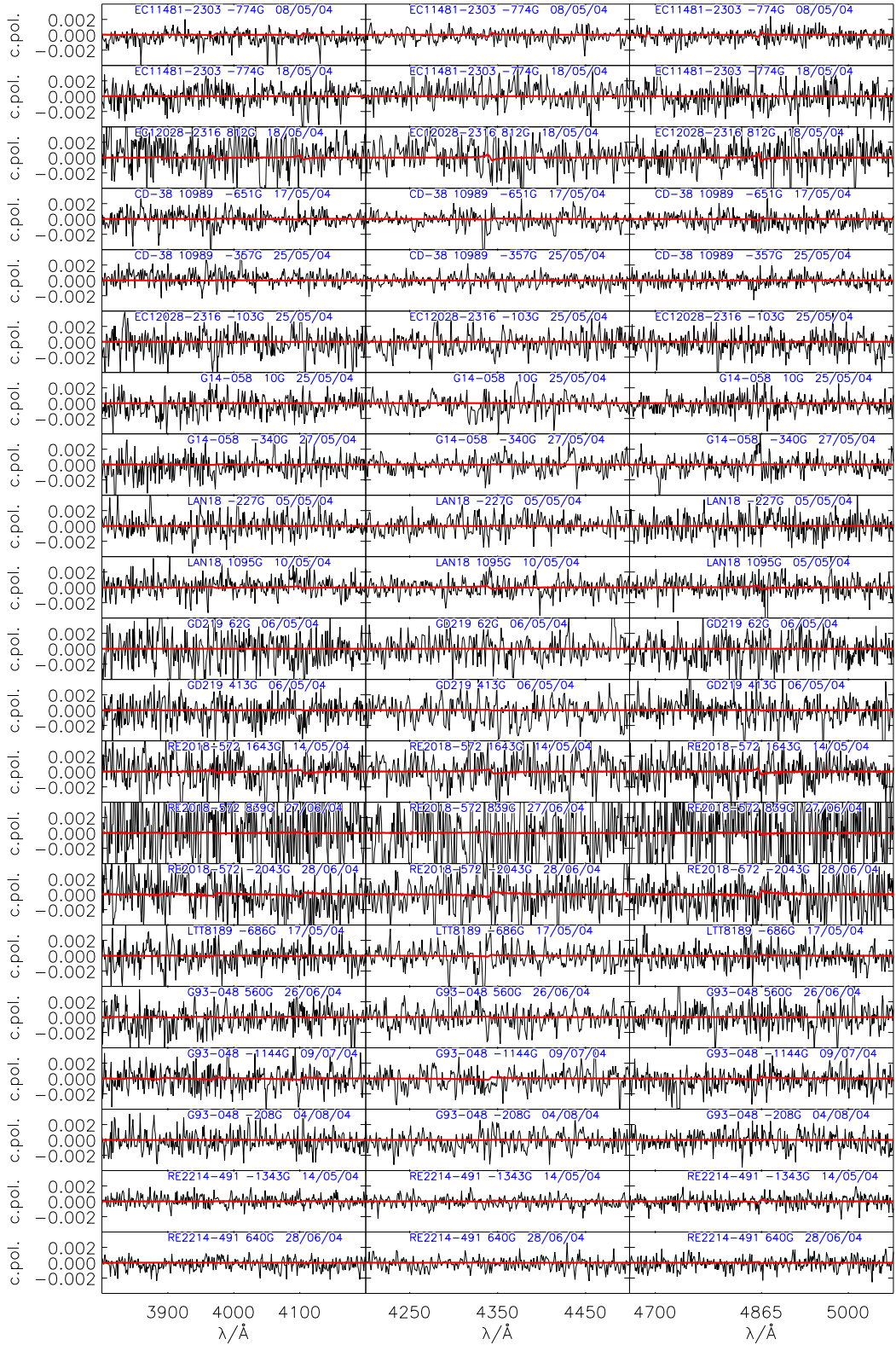}
   \caption{Circular polarisation spectra $V/I$ (thin black lines)
 for different observing dates in the regions 
around H$\beta$ (4861\AA, right), H$\gamma$ (4340\,\AA, middle), and close
to the Balmer threshold (left) for those measurements, where no significant
magnetic field could be detected. The red thick lines
 represent the circular 
polarisation predicted by the low-field approximation (Eq.\,\ref{e:lf}) using
the values for the longitudinal magnetic field from the best fit to the
H$\beta$ and  H$\gamma$ regions. }
              \label{wd73a2}%
    \end{figure*}
   \begin{figure*}
   \centering
   \includegraphics[width=0.78\textwidth,angle=270]{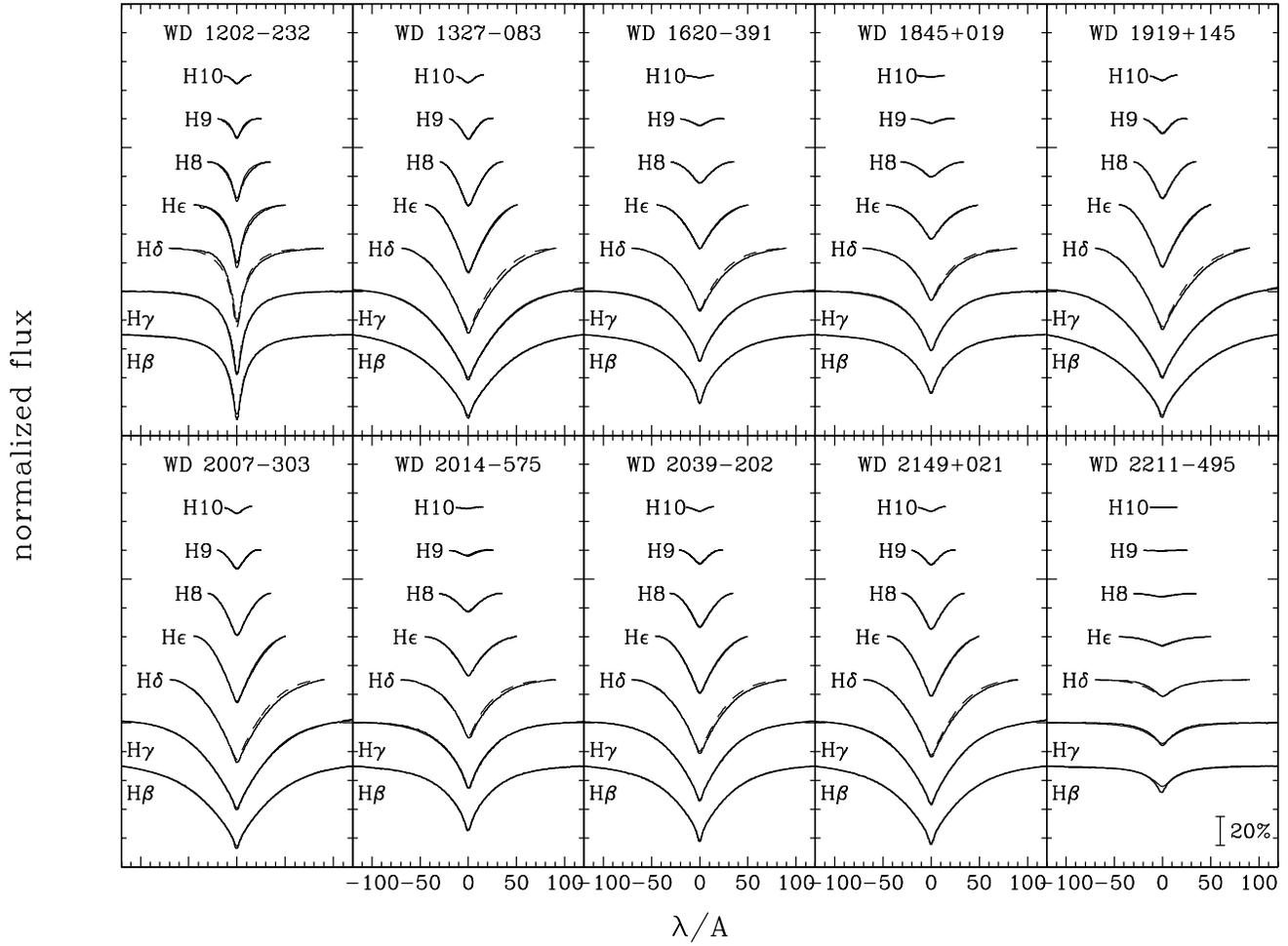}
   \caption{Model atmosphere fits (solid line) for the mean
   of all measured spectra (dashed line) of the white dwarfs. The results for 
$T_{\rm eff}$ and $\log g$ are given in Table\,\ref{t:lit}.}
              \label{f:spectra}%
    \end{figure*}

\end{appendix}

\begin{thebibliography}{}
\bibitem[\protect\citeauthoryear{Angel \& Landstreet}{1970}]{Angel-Landstreet:70}
         Angel J. R. P., Landstreet J. D., 1970, ApJ 160, L147
\bibitem[\protect\citeauthoryear{Aznar Cuadrado et al.}{2004}]{Aznar-etal:04}
Aznar Cuadrado R., Jordan S., Napiwotzki R., Schmid H.M.,
Solanki S.K., Mathys G., 2004, A\&A 423, 1081
\bibitem[\protect\citeauthoryear{Bagnulo et al.}{2006}]{Bagnulo-etal:06}
          {Bagnulo S., Landstreet J.D., Mason E.,
          Andretta V., Silaj J., Wade G.A., 
          2006, A\&A 450, 777}
\bibitem[\protect\citeauthoryear{Benvenuto \& Althaus}{1999}]{Benvenuto+Althaus:99}
          Benvenuto O.G., Althaus, L.G. 1999, MNRAS 303, 30
\bibitem[\protect\citeauthoryear{Berger et al.}{2005}]{Berger-etal:05}
         Berger, L., Koester, D., Napiwotzki R., Reid, I.N., Zuckermann, B., 2005, A\&A 565, 571 
\bibitem[\protect\citeauthoryear{Bragaglia et al.}{1995}]{Bragaglia-etal:95}
         Bragaglia A., Renzini A., Bergeron P., 1995, ApJ 443, 735
\bibitem[\protect\citeauthoryear{Casini \& Landi degl'Innocenti}{1994}]{Casini-Landi:94}
         Casini R., Landi degl'Innocenti E., 1994, A\&A 291, 668
\bibitem[\protect\citeauthoryear{Fabrika et al.}{2003}]{Fabrika-etal:03}
         Fabrika S. N., Valyavin G. G., Burlakova T. E., 2003,
         Astronomy Letters 29, 737
\bibitem[\protect\citeauthoryear{Finley et al.}{1997}]{Finley-etal:97}
         Finley D. S., Koester D., Basri G., 1997, ApJ 488, 375
\bibitem[\protect\citeauthoryear{Jordan et al.}{2005}]{jordanetal05-1}
{Jordan}, S., {Werner}, K., \& {O'Toole}, 2005, A\&A, 432, 273
\bibitem[\protect\citeauthoryear{Karl et al.}{2005}]{Karl-etal:05}
         Karl C.A., Napiwotzki R., Heber U., Dreizler S., Koester D., Reid I.N., 2005, A\&A 434, 637
\bibitem[\protect\citeauthoryear{Kilkenny et al.}{1997}]{Kilkenny-etal:97}
          Kilkenny, D., O'Donoghue, D., Koen, C., Stobie, R. S., Chen, A. 1997,
MNRAS 287, 867
\bibitem[\protect\citeauthoryear{Koester et al.}{1998}]{Koester-etal:98}
         Koester D., Dreizler S., Weidemann V., Allard N. F., 1998,
         A\&A 338, 612
\bibitem[\protect\citeauthoryear{Koester et al.}{2001}]{Koester-etal:01}
         Koester D., Napiwotzki R., Christlieb N., et al. 2001, A\&A 378, 556
\bibitem[\protect\citeauthoryear{Landi degl'Innocenti \& Landi degl'Innocenti}{1973}]{Landi-Landi:73}
         Landi degl'Innocenti E., Landi degl'Innocenti M., 1973,
         Solar Phys. 29, 287
\bibitem[\protect\citeauthoryear{Landstreet}{1982}]{Landstreet:82}
         Landstreet J. D., 1982, ApJ 258, 639
\bibitem[\protect\citeauthoryear{Liebert}{1988}]{Liebert:88}
         Liebert J., 1988, PASP 100, 1302
\bibitem[\protect\citeauthoryear{Liebert et~al.}{2003}]{liebertetal03-1}
{Liebert}, J., {Bergeron}, P., \& {Holberg}, J.~B. 2003, AJ, 125, 348
\bibitem[\protect\citeauthoryear{McCook \& Sion}{1999}]{McCook-Sion:99}
         McCook G. P., Sion E. M., 1999, ApJS 121, 1
\bibitem[\protect\citeauthoryear{Napiwotzki et al.}{2003}]{Napi-etal:03}
         Napiwotzki R., Christlieb N., Drechsel H., et al., 2003,
         ESO-Messenger 112, 25
\bibitem[\protect\citeauthoryear{O'Toole et~al.}{2005}]{otooleetal05-2}
{O'Toole}, S.~J., {Jordan}, S., {Friedrich}, S., \& {Heber}, U. 2005, A\&A,
  437, 227
\bibitem[\protect\citeauthoryear{Perryman et al.}{1997}]{Perryman-etal:97}
         Perryman, M.A.C, Lindegren, L., Kovalevsky J., et al., 1997,
          A\&A 323, L49
\bibitem[\protect\citeauthoryear{Schmidt et~al.}{2003}]{schmidtetal03-1}
{Schmidt}, G.~D., {Harris}, H.~C., {Liebert}, J., {et~al.} 2003, ApJ, 595, 1101
\bibitem[\protect\citeauthoryear{Stys et al.}{2000}]{Stys-etal}
         Stys, D., Slevinsky, R., Sion, E.M., Saffer, R., PASP 112, 354
\bibitem[\protect\citeauthoryear{Valyavin et al.}{2006}]{Valyavin-etal:06}
         Valyavin G., Bagnulo S., Fabrika S., Reisenegger A., Wade G.A.,
         Han I., Monin D., 2006, ApJ in press, astro-ph/0605401
\bibitem[\protect\citeauthoryear{van Altena et al.}{1995}]{vanAltena-etal:95}
        van Altena, W.\ F., Lee, J.\ T., Hoffleit, E.\ D. 1995, The general
        catalogue of trigonometric parallaxes, $4^\mathrm{th}$ ed.,
        (New Haven: Yale University Observatory)
\bibitem[\protect\citeauthoryear{Vanlandingham et~al.}{2005}]{vanlandinghametal05-1}
{Vanlandingham}, K.~M., {Schmidt}, G.~D., {Eisenstein}, D.~J., {et~al.} 2005,
  AJ, 130, 734
\bibitem[\protect\citeauthoryear{Werner}{1986}]{Werner:86}
         Werner K., 1986, A\&A 161, 177
\bibitem[\protect\citeauthoryear{Wickramasinghe \& Ferrario}{2000}]{wickramasinghe+ferrario00-1}
{Wickramasinghe}, D.~T. \& {Ferrario}, L. 2000, PASP, 112, 873
\bibitem[\protect\citeauthoryear{Wickramasinghe \& Ferrario}{2005}]{wickramasinghe+ferrario05-1}
{Wickramasinghe}, D.~T. \& {Ferrario}, L. 2005, MNRAS, 356, 1576
\end{thebibliography}
\end{document}